\documentclass[aps,prl,twocolumn,showpacs,amsmath,amssymb,letterpaper, floatfix, 10pt]{revtex4-1}
\usepackage[pdftex]{graphicx} 
\usepackage{amsmath}
\usepackage{amssymb}
\usepackage{wasysym}
\usepackage[
pdfstartview=XYZ,
bookmarks=false,
colorlinks=true,
linkcolor=blue,
urlcolor=blue,
citecolor=blue,
pdftex,
bookmarks=false,
linktocpage=true, 
hyperindex=false
]{hyperref}
\usepackage{psfrag}
\usepackage{CJKutf8}
\usepackage{tikz}

\usepackage[normalem]{ulem}

\begin{document}
\title{Interpreting current-induced spin polarization in topological insulator surface states}

\author{Pengke Li (\begin{CJK*}{UTF8}{gbsn}李鹏科\end{CJK*})}
\author{Ian Appelbaum}
\email{appelbaum@physics.umd.edu}
\affiliation{Department of Physics, Center for Nanophysics and Advanced Materials, U. Maryland, College Park, MD 20742}

\begin{abstract}
Several recent experiments on three-dimensional topological insulators claim to observe a large charge current-induced non-equilibrium ensemble spin polarization of electrons in the helical surface state. 
We present a comprehensive criticism of such claims, using both theory and experiment: First, we clarify the interpretation of quantities extracted from these measurements by deriving standard expressions from a Boltzmann transport equation approach in the relaxation-time approximation at zero and finite temperature to emphasize our assertion that, despite high in-plane spin projection, obtainable current-induced ensemble spin polarization is minuscule. Secondly, we use a simple experiment to demonstrate that magnetic field-dependent open-circuit voltage hysteresis (identical to those attributed to current-induced spin polarization in topological insulator surface states) can be generated in analogous devices where current is driven through thin films of a topologically-trivial metal. This result {\em{ipso facto}} discredits the naive interpretation of previous experiments with TIs, which were used to claim observation of helicity, i.e. spin-momentum locking in the topologically-protected surface state.
\end{abstract}
\maketitle

The prospect of electrically generating spin-polarized charge carriers without using ferromagnetic metals has intrigued the spintronics community for several years.\cite{Awschalom_Physics09} Possible schemes to achieve this end include spin-Hall effect\cite{Dyakonov_PLA71, Valenzuela_nature2006} and current-induced spin polarization\cite{Edelstein_SSC90, Kato_PRL04} in materials with strong spin-orbit coupling. 

\begin{figure}
\includegraphics[scale=0.85]{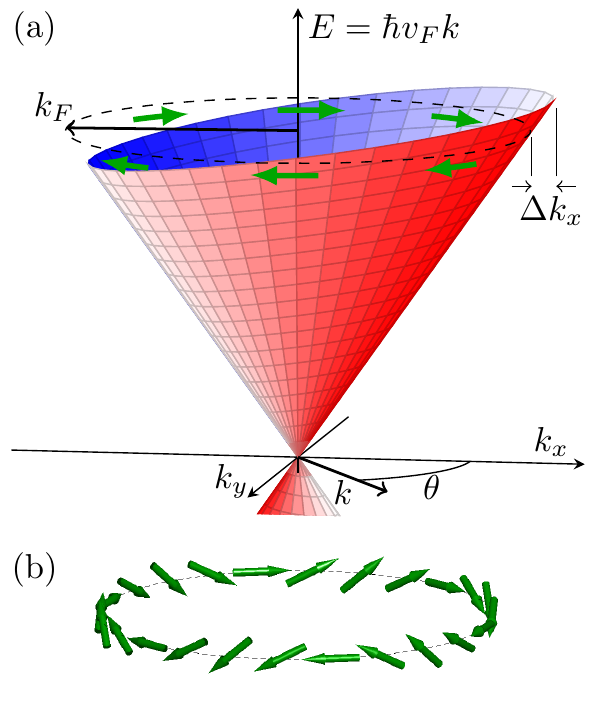}
\vspace{-10pt}
\caption{(a) Spin-polarized dispersion of the gapless surface states in a generic three-dimensional topological insulator, showing out-of-equilibrium occupied states at zero temperature in an electric field along $x$. Outer surface shading in red is proportional to spin-up component, inner shading in blue is proportional to spin-down in a $y$-axis spin basis. The equilibrium Fermi surface is shown as a dashed circle parallel to the $k_x-k_y$ plane, with ideal spin orientation shown in green arrows. (b) In real TI systems, the in-plane spin projection is reduced by out-of-plane canting, leading to a non-unity polarization $P_{SS}$.  \label{fig:dispersion}}
\end{figure}

The recent discovery of topological insulators (TIs) has fueled interest in attempts to meet this challenge. In these TI materials, spin-orbit interaction is so strong that it leads to an inverted bandgap, necessitating the existence of gapless 2-dimensional states at the isolated surface of a three-dimensional bulk. As shown in Fig. \ref{fig:dispersion}(a), these states have a linear dispersion $E=\pm\hbar k v_F$ near the degeneracy (or `Dirac') point and ideally have a helical spin orientation $\vec k \times \hat z$. This rigid relationship between $\vec k$ and spin is often dubbed ``spin-momentum locking".

At equilibrium, the opposite group velocities for anti-parallel spins in this simple model has been said to result in a perfectly-polarized (and ``dissipationless") ``spin current".\cite{Murakami_Science03} However, the relevance of this putative spin current to the goals of realizing useful devices is questionable at best, since it is no more physical than the equally ``dissipationless" surface charge current on a magnetized solid.\cite{Resta_JPhysC10} Neither can be coupled to other materials to do useful work. 

In addition to this sterile surface spin current polarization, deviations from the ideal in-plane spin orientation result from out-of-plane contributions to the effective spin-orbit fields, lowering their intrinsic state ``polarization".\cite{Yazyev_PRL10} For example, threefold rotational symmetry in Bi$_2$Se$_3$ leads to a hexagonal warping\cite{Fu_PRL09} due to a spin-orbit coupling similar to Dresselhaus spin splitting in some 2-dimensional semiconductors with the same rotational symmetry.\cite{Li_PRB15} This non-ideal spin texture is schematically illustrated in Fig. \ref{fig:dispersion}(b).

Either of these spin ``polarizations" can easily be confused with the \emph{density} spin polarization that accompanies momentum asymmetry, induced by driving a real charge current through the occupied surface states.\cite{Culcer_PRB10, Burkov_PRL10} Unlike the surface spin current, this spin polarization may conceivably be coupled to an external material for spintronic purposes.\cite{Appelbaum_APL11} 

The possibility of using the current-induced spin polarization in the surface states has motivated many recent experiments. By measuring the open-circuit voltage on a ferromagnetic contact\cite{Johnson_PRL85}, claims of surface state spin polarizations in the range 20-80\% have been made.\cite{Li_NNano14, Dankert_NanoLett15, Tian_SciRep15, Lee_PRB15, Ando_NanoLett14} If this is indeed an ensemble spin density polarization induced by the flow of current (whether from surface or bulk states), these values would establish TIs as a viable means for producing spin-polarized electrons without ferromagnetic elements, so long as the power dissipation from the necessary current flow is tolerated.\cite{Tang_NanoLett14}

Which spin polarization has been measured (if at all)? The current-induced spin polarization (a property of the nonequilibrium ensemble) or the intrinsic in-plane spin projection of the states themselves?\cite{Yazyev_PRL10, Fu_PRL09} Due to the accumulation of similar experiments reported in the literature, and prevalence of ambiguous spin polarization claims associated with this method, we find it essential to examine a realistic and well-supported model. The purpose of the present manuscript is therefore to provide both an explanation of the elementary theory, and a simple experiment on a trivial conductor which together clarify the meaning and veracity of previous claims. We show that even if the in-plane spin projection (the intrinsic spin ``polarization") is of the order of unity, the maximal ensemble spin polarization is minuscule. Our experiments with topologically trivial Au thin films indicate that even if driven charge currents have indeed induced spin polarization, spin-momentum locking in a topologically non-trivial surface state is by no means a necessary condition.   

{\em{Total current--}} A charge current density $j$ induced by an electric field $\mathcal{E}_x$ distorts the zero-temperature occupation function $f(\vec{k})=f_0(\vec{k})+g(\vec{k})$ out of equilibrium, where $f_0(\vec k)$ is the spherically-symmetric zero-temperature Fermi-Dirac distribution. Here, the asymmetric component $g(\vec{k})=\Delta k_x\frac{df}{dk_x}$ (where $\Delta k_x=\frac{e\tau\mathcal{E}_x}{\hbar}$ is determined by algebraic solution of the linear Boltzmann equation) is responsible for the nonzero current density $j$, where $\tau$ is the momentum scattering time. We can calculate it via
\begin{equation}
j=-e\int v_x g(\vec{k})\frac{d^2k}{(2\pi)^2}=-e\int v_F\cos{\theta}\Delta k_x\frac{df}{dk_x}\frac{d^2k}{(2\pi)^2}. \notag 
\end{equation}
Since $\frac{df}{dk_x}=\frac{df}{dk}\frac{dk}{dk_x}=-\delta(k-k_F)\frac{k_x}{k}$ at zero temperature and $k_x=k\cos\theta$, we have 
\begin{align}
j=\frac{ek_Fv_F\Delta k_x}{(2\pi)^2}\int_0^{2\pi}\cos^2\theta d\theta=\left(\frac{e v_F k_F}{4\pi}\right)\Delta k_x.
\label{eq:j}
\end{align}

{\em{Spin density--}} As schematically shown by red and blue shading in Fig. \ref{fig:dispersion}, the spin wavefunction of a given TI surface state in the $y$-basis, assuming the simplest two-band model with 100\% in-plane spin polarization, is 
\begin{align}
|\chi\rangle=\cos\frac{\theta}{2}|\uparrow\rangle+\sin\frac{\theta}{2}|\downarrow\rangle.\notag
\end{align}
To calculate the spin density, we can thus sum the probability $|\langle \chi|\uparrow\rangle|^2$ of finding each of these states with spin-up  
\begin{align}
n_\uparrow
=&\int_0^{2\pi}\int_0^\infty \left(f_0(\vec k)+\Delta k_x\delta(k-k_F)\cos\theta\right) \cos^2\frac{\theta}{2}\frac{kdkd\theta}{(2\pi)^2}\notag\\
=&\frac{1}{8\pi}(k_F^2+k_F\Delta k_x).
\notag
\end{align}
The result for spin-down density is similarly $n_\downarrow=\frac{1}{8\pi}(k_F^2-k_F\Delta k_x)$. The current-induced spin density polarization is thus 
\begin{equation}
P_{CI}=\frac{n_\uparrow-n_\downarrow}{n_\uparrow+n_\downarrow}
=\frac{\Delta k_x}{k_F}.
\label{eq:P}
\end{equation}
The Fermi wavevector $k_F$ is set by the equilibrium charge density, whereas  $\Delta k_x$ is determined by Eqn. \ref{eq:j}. This allows us to write
\begin{equation}
P_{CI}=\frac{j}{j_0}, \text{ where } j_0=\frac{e}{4\pi}v_Fk_F^2.
\label{eq:Pj}
\end{equation}
\noindent 
Note that in terms of the drift velocity $v_d$, $j=(n_\uparrow+n_\downarrow)ev_d=\frac{ev_d k_F^2}{4\pi}$. Thus we can equivalently write $P_{CI}=\frac{v_d}{v_F}$, consistent with the assertion in item (4) on p. 3 of Ref \onlinecite{Appelbaum_APL11}.

Already from Eqn. \ref{eq:P} we can see that Bi$_2$Se$_3$ is particularly disadvantaged for generating current-induced spin polarization. Even if electron-(acoustic)phonon coupling is weak, strong coupling to an optical phonon with energy $E_{op}=8$~meV \cite{Butch_PRB10} will limit $\Delta k_x$. Attempts to force the system beyond this bottleneck with ever larger electric fields will only cause energy dissipation via Joule heating which leads to the creation of thermal bulk carriers that dilute the contribution of the surface state to the net current. Thus, with a metallic sample having a chemical potential $\epsilon_F$ in the conduction band, the upper limit for spin polarization is $ E_{op}/\epsilon_F \apprle$ 8meV/300meV $\approx$ 3\%.

{\em{Open circuit voltage--}} Transport experiments designed 
to measure the surface state spin polarization ultimately must relate it to the open-circuit voltage $V$ of a ferromagnetic metal (FM) in contact with the TI across a tunnel barrier. This voltage is present to maintain zero net current through the interface, subject to a $\vec k$-dependent conductance $G_I=G_\uparrow\cos^2\frac{\theta}{2}+G_\downarrow\sin^2\frac{\theta}{2}$.

By writing the interfacial current using Ohm's law
\begin{align}
\int_0^{2\pi}G_I \left(
\Delta V- \frac{\hbar v_F}{e}\Delta k_x \cos\theta \right)d\theta=0,
\end{align}
we can solve for the induced open-circuit voltage
\begin{equation}
\Delta V=\frac{P_{fb}\hbar v_F}{2e}\Delta k_x, \label{eq:DV}
\end{equation}
\noindent where $P_{fb}=\frac{G_\uparrow-G_\downarrow}{G_\uparrow+G_\downarrow}$ is the ``effective spin polarization of the ferromagnetic-barrier couple" \cite{Slonczewski_PRB89}, which approaches the bulk ferromagnet spin polarization $P_{FM}$ in the thick barrier limit. Using Eqn. \ref{eq:P}, we can then re-write Eqn. \ref{eq:DV} as
\begin{align}
\Delta V=\frac{h}{e^2}\left(\frac{ev_Fk_F}{4\pi}\right)P_{fb}P_{CI}=\frac{h}{e^2}\frac{j}{k_F}P_{fb}.
\label{eq:DV2}
\end{align}
Note that the quantity in parenthesis $\frac{ev_Fk_F}{4\pi}=\frac{j_0}{k_F}$ has units of a fundamental charge current unrelated to the surface current density $j$. Here, we have assumed perfect spin polarization of the surface states themselves (i.e. states with opposite $\vec k$ can be expressed as spin up and down); accounting for deviations from this condition, as when considering a model with more than just two bands,\cite{Yazyev_PRL10, Fu_PRL09, Chang_PRB15} involves inclusion of an intrinsic surface state spin polarization factor $P_{SS}$ in Eqn. \ref{eq:DV2}. This quantity is a measure of the average in-plane spin projection of states near the chemical potential, as shown in Fig. \ref{fig:dispersion}(b). 

Other than an errant factor of $\pi$ and approximation of $P_{fb}$ with $P_{FM}$, our result Eqn. \ref{eq:DV2} is consistent with the first equation of Ref. \onlinecite{Hong_PRB12}, which is often used in the analysis of open-circuit voltage spin detection experiments on TIs. In that paper, the authors assert an expression equivalent to 
\begin{align}
\Delta V=\frac{h}{e^2}\left(\frac{\pi j}{k_F}\right)P_{FM}P_{SS}.\qquad\text{[Ref. \onlinecite{Hong_PRB12}, Eqn. 1]}\label{eq:Chen}
\end{align}
where we simply use our variable definitions and $j=I/W$, where $I$ is total charge current and $W$ is the width of the FM/TI interface transverse to the current flow. This equation is also consistent up to a factor of $2$ with the expression given in Ref. \onlinecite{deVries_PRB15}. Any such trivial deviations with prior results from others' works are irrelevant to the main message of the theoretical part of this paper, which follows.

The surface state polarization (i.e. in-plane spin projection) $P_{SS}$ extracted from measurements of $\Delta V$ in this way have values of 0.2-0.8.\cite{Li_NNano14, Dankert_NanoLett15, Tian_SciRep15, Lee_PRB15}  However, one must remember that \emph{this is not the current-induced spin polarization} $P_{CI}$ given in Eqn. \ref{eq:Pj}, which is typically two orders of magnitude smaller. As we calculate below in the Discussion, it is far too small to establish the viability of ``\dots using TIs as spin polarized sources for spintronic devices at ambient temperatures"\cite{Dankert_NanoLett15}.  

{\em{Finite Temperature-- }}Our zero temperature model is limited to the linear response regime where we expect ohmic transport. Eqn. \ref{eq:Pj} is still correct at finite temperature $T$, but the current density (Eqn. \ref{eq:j}) and the current-induced spin polarization (Eqn. \ref{eq:P}) are modified by the necessity to integrate over the gradient of Fermi-Dirac distribution $f$, yielding a quantity
\begin{align}
\int_0^\infty \frac{df(k,\beta,\epsilon_F)}{dk}k dk=-\frac{1}{\hbar v_F\beta}\ln(1+e^{\beta\epsilon_F}),\notag
\end{align}
where $\beta=1/k_BT$ and $\epsilon_F$ is the chemical potential. Then,
\begin{align}
j=&\frac{e \Delta k_x}{4\pi\hbar\beta}\ln(1+e^{\beta\epsilon_F}),\\
P_{CI}=&\frac{\ln(1+e^{\beta\epsilon_F})}{\beta\epsilon_F}\frac{\hbar v_F\Delta k_x}{\epsilon_F}.\label{eq:PT}
\end{align}
The appropriate zero-temperature expressions are clearly recovered in the $\beta\epsilon_F\rightarrow\infty$ limit. Furthermore, at non-zero temperatures the current-induced spin polarization is slightly enhanced: for $\beta\epsilon_F=1$, $P_{CI}$ from Eqn. \ref{eq:PT} is $\approx$30\% larger than the zero-temperature result (Eqn. \ref{eq:P}) at the same value of $\Delta k_x$. Note that this may require substantial increase in the electric field $\mathcal{E}_x$ since the relaxation time $\tau$ is likely to be  shortened by additional electron-phonon scattering. At room temperature, $k_BT\approx$30meV and the deviation from the zero-temperature result is negligible in Bi$_2$Se$_3$ with typical $\epsilon_F\approx 300$meV, as can be seen from expansion of $\frac{1}{x}\ln (1+e^x)\approx 1+e^{-x}/x$ in the limit of large $x$.

{\em{Discussion and Experiment}-- } With a chemical potential $\epsilon_F=\hbar v_Fk_F=100$~meV and $v_F\approx 10^8$~cm/s\cite{Zhang_NatPhys09}, Eqn. \ref{eq:Pj} gives $j_0\approx 3$~A/cm. Therefore, to drive a current-induced spin polarization of $\approx$1\%, one must source a 2D current density of $\approx 3\mu$A/$\mu$m.  With a mobility of $\mu=$1000~cm$^2$/Vs\cite{Butch_PRB10} and charge density of $10^{13}$cm$^{-2}$, the TI has resistivity $\rho=(nq\mu)^{-1}\approx$1 k$\Omega \Box$.   The Joule power is then $j^2\rho\approx $1W/cm$^2$ for 1\% polarization. Likewise, 10\% polarization demands $\approx $100W/cm$^2$, which is certain to induce substantial heating. The bulk gap in many TIs is only several hundred meV or less, and hence they likely cannot be used in this regime without an intolerably high contribution from bulk conduction due to interband thermal excitation.

\begin{figure}
 {\includegraphics[scale=0.425]{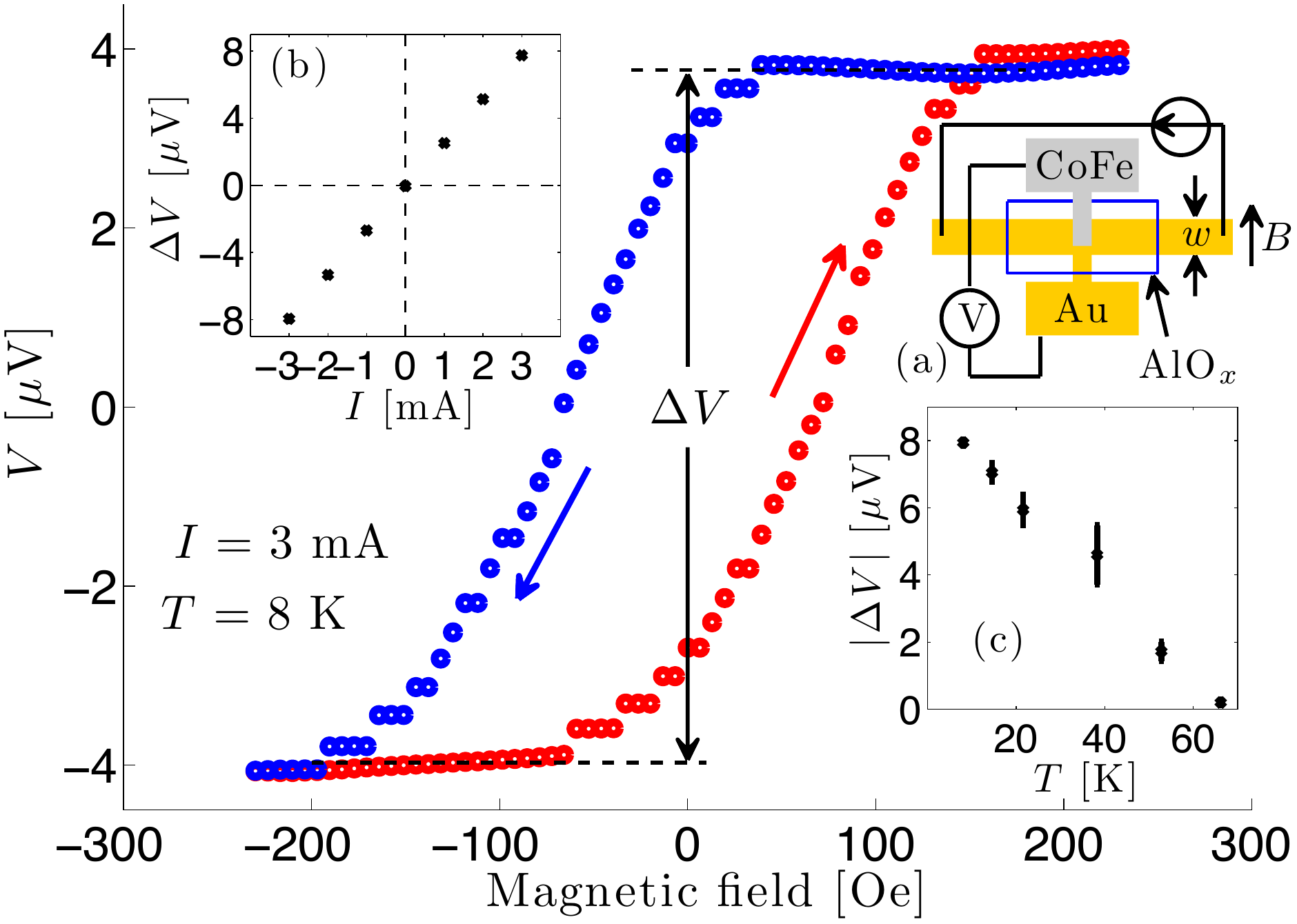}}
 \vspace{-7pt}
\caption{Nonlocal voltage hysteresis under transverse magnetic field following the measurement scheme given by Inset (a), where the Au transport channel width is $w=100\mu$m. The blue and red data corresponds to opposite field scanning directions, collected under $I=3$ mA current bias at $T$=8~K. Inset (b) shows the linear dependence of $\Delta V$ and $I$. Inset (c) shows that the amplitude of $\Delta V$ decreases monotonically as temperature increases.
\label{fig:expt}}
\end{figure}
 These parallel conducting channels from bulk-related carriers (either thermally generated or present from degenerate doping or interfacial band-bending) may or may not dilute the surface spin accumulation. The bulk spin Hall effect \cite{Dyakonov_PLA71,Valenzuela_nature2006} and Edelstein effect \cite{Edelstein_SSC90}, or ferromagnetic proximity can generate spin-polarization even without a helical (spin-momentum locked) surface state. To emphasize this point, we performed experiments on devices with ferromagnetic contact to the surface of a topologically-trivial thin Au film carrying $\approx 10^5$~A/cm$^2$ volume current density, using the same open-circuit voltage probe scheme employed in recent experiments on TIs and shown in Inset (a) to Fig. \ref{fig:expt}.\cite{Li_NNano14, Dankert_NanoLett15, Tian_SciRep15, Lee_PRB15} The 10~nm-thick Au film was thermally evaporated onto an oxidized Si wafer, followed by atomic layer deposition of 1 nm AlO$x$ and e-beam evaporation of 20 nm CoFe spin detector contacts. The metal films are patterned by shadow masks and all three steps are done \textit{ex situ}. Typical behavior of the open-circuit voltage used to sense surface spin accumulation during a magnetic field sweep at 8K results in the hysteresis seen in the main panel of Fig. \ref{fig:expt}. The $\Delta V\approx 8\mu$V change we observe is indistinguishable in magnitude from analogous measurements on TIs such as Bi$_2$Se$_3$, and appropriately scales with current magnitude and direction as shown in Inset (b). Although its origin is unclear, we can rule out spurious contributions from e.g. anisotropic magnetoresistance because it has a different symmetry, and anomalous Hall effect because it is strongly dependent on temperature (similar to measurements with disordered thin-film Bi$_2$Se$_3$, Ref. \onlinecite{Li_NNano14}), as indicated in Inset (c). This observation of nominally identical measurement signals using a trivial metal casts serious doubts on the physical interpretation of existing experiments with TIs.

{\em{Conclusion}-- } We have shown that the current-induced polarization in 2-dimensional topological surface states is, in general, much smaller than the polarization of conventional spin sources, i.e. conduction electrons in a typical metallic ferromagnet, which can easily be several tens of percent. However, our derivation ignores the further possibility that the measured open-circuit voltage signal is polluted by spurious effects and its physical interpretation is ambiguous to begin with.\cite{deVries_PRB15} Our experimental results on devices fabricated with topologically-trivial metallic Au thin films reproduce nominally identical open-circuit voltage hysteresis seen with topological insulators, making that possibility entirely plausible. This obvious control experiment 
shows that there can be many deceptive sources of the magnetization-dependent open circuit voltage besides a spin-polarized topologically-protected surface state. To resolve the true origin of these signals, further work with alternative transport methods is required, and constitutes a topic entirely separate from specious issues of topology.

 Even assuming that signals due to surface-state spin polarization can be disentangled from bulk spin Hall, Edelstein and proximity effects, we must conclude that the only apparent way to circumvent the essential limitation imposed by Eqn. \ref{eq:Pj} is to approach the charge-neutrality condition when chemical potential is close to the Dirac point. However, the total charge density is then simultaneously minimized, so that the TI surface can not provide many carriers regardless of their spin polarization. Ultimately, one struggles to justify the cost of high collateral power dissipation in a quest to generate a meager spin polarization.  This is especially true when the generation mechanism itself remains ambiguous -- as highlighted by our  experiments showing identical measurements on completely trivial materials.

{\em{Acknowledgment-- }}This work was supported by the Office of Naval Research under contract N000141410317, and the Defense Threat Reduction Agency under contract HDTRA1-13-1-0013.

\bibliography{TI}

\end{document}